\documentstyle[onecolumn]{mn}
\def\psfig#1{}
\def\df{{\sc df}}
\def\half{\textstyle{\frac12}}
\def\erf{\mathop{\rm erf}\nolimits} 
\begin{document}

\title[Galaxy Models]{Nearly Self-Consistent Disk-Bulge-Halo Models for
Galaxies}

\author[Kuijken and Dubinski]{Konrad Kuijken$^1$ and John
Dubinski$^2$\\
$^1$ Kapteyn Institute, PO Box 800, 9700 AV Groningen, The
Netherlands\\
$^2$ Lick Observatory, University of California, Santa Cruz, CA
95064, USA}

\maketitle

\begin{abstract}
We describe methods for setting up self-consistent disk-bulge-halo galaxy
models.  The bulge and halo distribution functions (\df) are functions
of $E$ and $L_z$ only.  The halo's flattening and rotation can be
specified.
The disk \df\ is a function of $E$ and $L_z$
and a third ``integral'', $E_z$, the vertical energy,
which is approximately conserved in a warm disk with vertical extent.
The models also have
finite extent making them suitable for N-body simulation.
A simulation of a sample model shows that in practice the models are very close
to equilibrium making them ideal for experiments on instabilities in
galactic disks.  We also present a sequence of models closely resembling
the Milky Way mass distribution with 5 exponential scale radii and varying
halo mass and radial extent.
\end{abstract}


\section{Introduction}
\noindent
This paper describes a set of semi-analytic models for the
phase-space distribution functions of axisymmetric disk galaxies. They
contain three components, corresponding to a disk, bulge and
halo. The distribution functions are simple functions of three
integrals of motion: the two analytic ones (energy and angular
momentum about the axis of symmetry) and one approximate integral
which describes the vertical motions in the disk component. These
distribution functions yield a unique density for each component
in any given potential: since all three components affect each
other gravitationally, a numerical solution of the Poisson
equation is required for self-consistent models.

Setting up a stable 3 dimensional
disk for N-body experiments is a difficult task.
Strategies to handle this problem have included adiabatically growing the
disk mass distribution in a self-consistent halo/bulge model (e.g. Barnes
1988), treating the the halo and bulge as a static background (e.g.
Sellwood \& Merritt 1994; Quinn, Hernquist \& Fullagar 1993), or directly
solving the Jeans equations (under a suitable Ansatz) for the complete
system of the disk, bulge and
halo to find the velocity dispersions (Hernquist
1993), and realizing these dispersions with (typically) Gaussian
distributions. In a recent study,
Sellwood \& Merritt (1994) used the \df\ calculated by Kalnajs (1976)
of the Kuzmin-Toomre disk to set up the radial and azimuthal
velocities, but this method is restricted to that class of models, and
they did use
the Jeans  equations for the vertical velocities.

While these methods are useful for disks with small $N$ where the
discreteness noise is dominant, as $N$ increases the approximate
initial conditions introduce subtle transient behaviour which can
interfere with the interpretation of experiments. Adiabatic growth of
the disk, or relaxation from non-equilibrium initial conditions have
the further disadvantage that the relaxed initial conditions are not
under perfect control of the experimenter. It is for example not
uncommon to observe outward propagating rings of overdensity from the
warmer disk center (where the Gaussian approximation is the poorest),
and transients in the disk velocity dispersion which can change the
initial values by $\sim 20$\% before the particles settle into
equilibrium (Hernquist, Mihos, and Walker, private communication).  As
a result, though these models relax quickly to equilibrium, the
velocity dispersion profiles change from the input values.  The
density profiles usually do not change significantly.  While such
subtle effects do not matter in experiments in which a disk will be
strongly perturbed (say by a merger), they can be significant in
investigations of the rather subtle instabilities which arise from the
inherent fragility of a cool disk.  Examples of these problems are
spiral structure formation (Sellwood \& Carlberg 1984), bar
instabilities (e.g. Hernquist and Weinberg 1992), bending
instabilities in counter-streaming disks (e.g. Sellwood \& Merritt
1994; Merritt \& Sellwood 1994), disk warping (Dubinski \& Kuijken
1995), and disk heating by satellite accretion or other tidal
perturbations (e.g. Quinn, Hernquist \& Fullager 1993).  Simulations
with $N$ in the millions are now becoming possible with greater
computing power.  The subtle transients at start-up due to various
approximations which were previously drowned out by discreteness,
become correspondingly more bothersome.  A cleaner set of initial
conditions are desireable.

In this paper we describe the construction of our models. In
\S\ref{sect-df}, we present the disk distribution function (\df) of
the disk, bulge and halo and a describe a method for calculating the
potential of the self-consistent model.  The bulge and halo have the
\df\ of King (1966) models and their flattened generalization, the
lowered Evans distributions (Kuijken \& Dubinski 1994). These \df s
depend only on the classical integrals of motion, and can be combined
straightforwardly.  The disk distribution function is a generalization
to three dimensions of the planar model devised by Shu (1969) (see also
Kuijken \& Tremaine 1992), similar to the construction by Binney
(1987), and allows almost arbitrary specification of the radial
variation of density and velocity dispersion. Since two-integral
distribution functions for disks are not realistic (Oort 1965), we
employ an approximate third integral.  \S\ref{sect-nbody} contains
N-body demonstrations and tests of equilibrium for a specific model.  In
\S\ref{sect-mw}, we present a sequence of models with different halo
extent suitable for describing the Milky Way. Finally, we discuss the
advantages and limitations of these models in \S\ref{sect-discuss}
and summarize our results.

\section{Distribution Function}
\label{sect-df}
\noindent
Our strategy is an extension of the technique we employed in the
construction of the lowered Evans models (Kuijken \& Dubinski 1994),
and earlier applied by e.g., Prendergast \& Tomer (1970) and Rowley
(1988).  The starting point is a chosen analytic form for the \df,
written in terms of known integrals of motion. Any such \df\
represents an equilibrium model in any gravitational potential which
respects these integrals, with space density explicitly determined by
this potential. We can therefore construct self-gravitating models by
requiring that the potential and the density be related by Poisson's
equation. This last step must generally be done numerically.

In order of increasing complexity, we now detail the \df's of the
three galactic components.

\subsection{The Bulge Distribution Function}
\noindent
For the bulge \df\ we take a King model (King 1966). This \df\
has the form
\begin{equation}
f_{\rm bulge}(E) =\cases{\rho_b (2\pi\sigma_b^2)^{-3/2}
\exp[(\Psi_o-\Psi_{c})/\sigma_b^2]
\{ \exp[-(E-\Psi_{c})/\sigma_b^2]-1 \} & if $E < \Psi_{c}$, \cr
                0 & otherwise.}
\label{eq-dfbulge}
\end{equation}
It depends on the three parameters $\Psi_c$ (the cutoff
potential of the bulge), $\rho_b$ (approximately the central
bulge density, ignoring the effects of the \df\ truncation) and
$\sigma_b$, which governs the velocity dispersion of the bulge
component. $\Psi_0$ is the gravitational potential at the center of
the model.

In what follows, we will normally choose $\sigma_b<\sigma_0$ and
$\Psi_c<0$ to make the bulge more centrally condensed, and more
radially confined, than the halo (the latter has a cutoff at zero
energy).

\subsection{The Halo Distribution Function}
\noindent
We use the \df\ of a lowered Evans model (Kuijken \& Dubinski
1994) for the halo. This \df\ is a truncation at finite energy
of the models discovered by Evans (1993) for the flattened
logarithmic potential. It takes the form
\begin{equation}
f_{\rm
halo}(E,L_z^2)=\cases{[(AL_z^2+B)\exp(-E/\sigma_0^2)+C][\exp(-E/\sigma_0^2)-1]
&
if $E
<0$, \cr
 0 & otherwise.}
\label{eq-dfhalo}
\end{equation}
The density corresponding to this \df\ is given by eq. 9 of
Kuijken \& Dubinski (1994), and is repeated here for
completeness:
\begin{eqnarray}
\rho_{\rm halo}(R,\Psi) &=&\half\pi^{3/2} \sigma_0^3(A R^2 \sigma_0^2 + 2B)
\erf(\sqrt{-2\Psi}/\sigma_0)\exp(-2\Psi/\sigma_0^2)  \nonumber \\
&&+(2\pi)^{3/2}\sigma_0^3(C-B-AR^2\sigma_0^2)
\erf(\sqrt{-\Psi}/\sigma_0)\exp(-\Psi/\sigma_0^2) \nonumber \\
&&+\pi\sqrt{-2\Psi}[\sigma_0^2(3A\sigma_0^2R^2+ 2B-4C)
+\frac43\Psi(2C-A\sigma_0^2R^2)],
\label{denshalo}
\end{eqnarray}
where $\erf(x)=2\pi^{-1/2}\int_0^x\exp(-t^2)\,dt$ is the usual error
function.  The halo \df\ has five free parameters: the potential well
depth $\Psi_0$, the velocity and density scales $\sigma_0$ and $\rho_1$,
the halo core radius $R_c$ and the flattening parameter $q$, (the last
three of these contained within the parameters $A$, $B,$ and $C$). For
convenience, we have defined a characteristic halo radius $R_a$ which
replaces the density scale $\rho_1$ of Kuijken \& Dubinski (1994):
\begin{equation}
\label{eq-radef}
R_a=\left(3\over2\pi G\rho_1\right)^{1/2}\sigma_0e^{\Psi_0/2\sigma_0^2};
\end{equation}
it is roughly the radius at which the halo rotation curve, if
continued at its $R=0$ slope, would reach the value $2^{1/2}\sigma_0$.

Arbitrary amounts of rotation can be added to the halo model by
splitting the \df\ into parts with positive and negative~$L_z$.

\subsection{The Disk Distribution Function}
\noindent
In the construction of a realistic three-integral disk distribution
function, the issue of a third integral cannot be evaded (as it was
for the bulge and halo components). For it is an observed fact that in
the solar neighbourhood (e.g., Wielen 1974) and in the
disks of other galaxies (Bottema 1993) the vertical and
radial dispersions are different, which is not possible in any \df\
that depends only on energy and angular momentum.  The simplest
approximate third integral in an axisymmetric disk system is the
energy in the vertical oscillations,
$E_z\equiv\Psi(R,z)-\Psi(R,0)+\half v_z^2$.  It is quite well
conserved along nearly circular orbits which have no large radial or
vertical excursions. We will use this quantity as third integral for
the disk \df\ in our models. More sophisticated possible integrals
are discussed by Kent and de Zeeuw (1991).

Armed with a third integral, the disk distribution function can
be constructed by extending the planar \df\ discussed by Shu~(1969)
and Kuijken \& Tremaine~(1992) into the vertical dimension, similar to
the \df\ constructed by Binney~(1987). Thus, we have
\begin{equation}
\label{eq-diskdf}
f_{\rm disk}(E_p,L_z,E_z)= {\Omega(R_c)\over(2\pi^3)^{1/2}\kappa(R_c)}
{\widetilde{\rho_d}(R_c)\over
           \widetilde{\sigma_R}^2(R_c)\widetilde{\sigma_z}(R_c)}
\exp\left[-{E_p-E_c(R_c)\over\widetilde{\sigma_R}^2(R_c)}
                      -{E_z\over\widetilde{\sigma_z}^2(R_c)}\right].
\end{equation}

Here, $E_p\equiv E-E_z$ is the energy in planar motions, $L_z$ is the
specific angular momentum about the axis of symmetry, $R_c$ and $E_c$
are the radius and energy of a circular orbit with angular momentum
$L_z$, and $\Omega$ and $\kappa$ are the circular and epicyclic
frequencies at radius $R_c$.
The density corresponding to this \df\ is obtained by integrating over
the three velocity components. The $v_R$ and $v_z$ integrals are
straightforward, leaving the $v_\phi$-integral:
\begin{eqnarray}
\rho_{\rm disk}(R,z)&=&\int_0^\infty\left\{
\left[dv_\phi\equiv dR_c\left(R_c\kappa(R_c)^2\over 2R\Omega(R_c)\right)\right]
{2\widetilde{\rho_d}(R_c)\Omega(R_c)\over
(2\pi)^{1/2}\widetilde{\sigma_R}(R_c)\kappa(R_c)}\right.\times\nonumber\\
&&\left.\exp\left[-{\Psi(R,0)-\Psi(R_c,0)\over\widetilde{\sigma_R}^2(R_c)}-
  \left({R_c^2\over R^2}-1\right){v_c^2(R_c)\over2\widetilde{\sigma_R}^2(R_c)}
  -{\Psi(R,z)-\Psi(R,0)\over\widetilde{\sigma_z}^2(R_c)}\right]\right\}.\qquad
\label{eq-diskdfint}
\end{eqnarray}

In the $z=0$ plane, this expression reduces to $\widetilde{\rho_d}(R)$
with fractional error $O(\widetilde{\sigma_R}^2/v_c^2)$, and to the
same order the radial velocity distribution is Gaussian with with
dispersion $\widetilde{\sigma_R}(R)$ (see Kuijken \& Tremaine
1992). The essence of the construction in eq.~\ref{eq-diskdf} is the
replacement of the radius $R$ (which is not an integral of motion) by
the epicyclic radius $R_c$ (which is a function of angular momentum,
and therefore is conserved along orbits). In warm disks, in which
excursions from circular orbits are small but not neglegible, this
parametrization still provides a good starting point for constructing
a \df\ with given radial density and velocity dispersion profiles. The
vertical structure of this disk is approximately isothermal, with the
scale height set by the vertical velocity dispersion
$\widetilde{\sigma_z}(R_c)$ and the vertical potential gradient.

Observed large disk galaxies have vertical light profiles similar to
the $\hbox{sech}^2(z)$-dependence expected for a vertically
isothermal, self-gravitating sheet (van der Kruit \& Searle 1981):
therefore, as long as the disk is mainly confined in $z$ by its own
gravity, the \df\ of equation
\ref{eq-diskdf} will be a reasonable model.

In any gravitational potential, we can adjust the `tilde' functions
$\widetilde\rho$, $\widetilde{\sigma_R}$ and $\widetilde{\sigma_z}$ to
the desired disk characteristics. In this paper, we arrange for the
disk density to be approximately radially exponential and truncated:
\begin{equation}
\label{eq-diskden}
\rho_{\rm disk}(R,z)=
{M_d\over 2\pi R_d^2 z_d}e^{-R/R_d}
\hbox{ erfc}\left({r-R_{\rm out}\over2^{1/2}\delta R_{\rm out}}\right)
\exp \left[ \ln \hbox{sech$^2$ (1)} \frac{\Psi_z(R,z)}{\Psi_z(R,z_d)} \right]
\end{equation}
Here
$M_d$ is a parameter which is close to the mass of the disk unless the
disk is severely truncated or the vertical structure is far from
sech$^2(z/z_d)$.
The vertical density of these disks is constructed to depend
exponentially on the vertical potential
$\Psi_z(R,z)\equiv\Psi(R,z)-\Psi(R,0)$, and to drop from the
mid-plane value by a factor $\hbox{sech}^2(1)$ at a height of
$z_d$, similar to the behaviour of a constant thickness isothermal
sheet.

Given a total potential for the model, we then set the disk tilde
functions in the disk \df\ as follows. In the limit of
very small velocity dispersions
these functions are the actual mid-plane density and
velocity dispersions.  We first choose the function
$\widetilde{\sigma_R}(R_c)$, approximately determining the radial
velocity dispersion in the disk. $\widetilde\rho$ and
$\widetilde{\sigma_z}$ are then iteratively adjusted so that the
density on the mid-plane and at height $z=z_d$ agree with those of
equation \ref{eq-diskden}. It turns out that, at least for the models
described in the remainder of this paper, this recipe yields a \df\
which has a space density close to that given by equation
\ref{eq-diskden}.

\subsection{Calculation of the Combined Potential}
\label{sec-poisson}
\noindent

The distribution functions for the various galaxy components all
imply a unique volume density in a given potential. To construct
a self-gravitating model, we need to find the potential
in which the combined density is also the one implied by
Poisson's equation, i.e.
\begin{equation}
\label{eq-poisson}
\nabla^2\Psi(R,z)=4\pi G [\rho_{\rm
disk}(R,\Psi,\Psi_z)+\rho_{\rm bulge}(\Psi)+\rho_{\rm
halo}(R,\Psi)].
\end{equation}
We solve equation \ref{eq-poisson} using a spherical harmonic
expansion, following Prendergast \& Tomer (1970), with two significant
modifications. First, we have found that the disk density obtained by
integrating the disk \df\ (eq.~\ref{eq-diskdf}) over all velocities is
close to the value given by eq.~\ref{eq-diskden} (this was, after all,
what the disk \df\ was designed to do). We therefore use this latter
expression in the solution of Poisson's equation: this change avoids
the single integral that would have to be calculated numerically each
time the disk density was needed. Second, and more importantly,
for realistically thin disks a spherical harmonic
expansion is not very efficient, since high-order terms must be taken
before a good approximation can be obtained. Moreover, zero-thickness
disks, which form a very regular limit physically, require a
prohibitive number of terms in the series. We therefore construct an
analytic potential which represents the high-frequency terms
correctly, and only fit harmonics to the residue.

A possible analytic `high harmonics' disk potential $\Psi^\dagger_{\rm
disk}$ is obtained by vertically integrating the disk density twice
(i.e. by solving Poisson's equation for the disk component ignoring
the radial gradient terms). Assuming that the vertical profile of the
disk is that of a self-gravitating, plane-parallel sheet, this
approach yields
\begin{equation}
\label{eq-baddiskpot}
\Psi^\dagger_{\rm disk}=4\pi G\rho_d(R)z_d^2\ln\,\hbox{cosh}(z/z_d)
\end{equation}
There are two reasons why this potential, as written, is not
suitable for removing the high harmonics. The first is that the
density corresponding to this potential, $\nabla^2\Psi^\dagger_{\rm
disk}/4\pi G$, does not converge to 0 at large distances from
the origin. The residual density can therefore not be fitted
conveniently with spherical harmonic coefficients. We resolve
this problem by replacing cylindrical radius $R$ by spherical
radius $r$ in equation \ref{eq-baddiskpot}: since the radial
density profile has a cutoff, the resulting potential and its
corresponding density will indeed converge at large radii, while
at small $z$ the potential still models well the higher frequency
terms of the disk. The second problem is that, even after
rewriting the equation in terms of $r$, the potential has discontinuous
derivatives at zero radius, if the disk density is radially
exponential. We therefore smoothly round off the potential in the
central disk scale length, by patching on a quartic function in
the central region. At small radii, the spherical harmonic
expansion is in any case better able to follow the disk density,
since the projected angle subtended by the disk, $z_d/R$, only
gets unmanageably small at large radii.

The high-frequency disk potential we use is then
\begin{equation}
\label{eq-hifreqdisk}
\Psi_{\rm disk}(r,z)={M_dz_d\over4\pi R_d^2}
\ln \hbox{cosh}(z/z_d) \times\cases{
\displaystyle\frac14e^{-1}[7-4(r/R_d)^2+(r/R_d)^4] & for $r<R_d$,\cr
\displaystyle\frac12
\hbox{ erfc}\left({r-R_{out}\over2^{1/2}\delta R_{out}}\right)
e^{-r/R_d}
& otherwise.\cr}
\end{equation}

The corresponding volume density is rather involved, but can be
obtained using the result
\begin{equation}
\nabla^2 f(r) \ln\,\hbox{cosh} z=f''(r) \ln\,\hbox{cosh}
z+2{f'(r)\over r}
(z\,\hbox{tanh z}+\ln\,\hbox{cosh}z)+f(r)\,\hbox{sech}^2z,
\end{equation}
where once again it should be remembered that $r$ is the spherical
radius. Note that the final term reproduces the disk density of
a sech$^2(z)$ disk to $O(z/R)^2$, and that the other terms have
traded a vertical derivative for a radial one, and so are
subdominant in thin disks. Note that it is important to have the
resolution to follow accurately the {\it radial} variation in
this density near the truncation radius, where the $f''$ term now
magnifies the change in disk density.

\begin{figure}
{\centerline{\psfig{figure=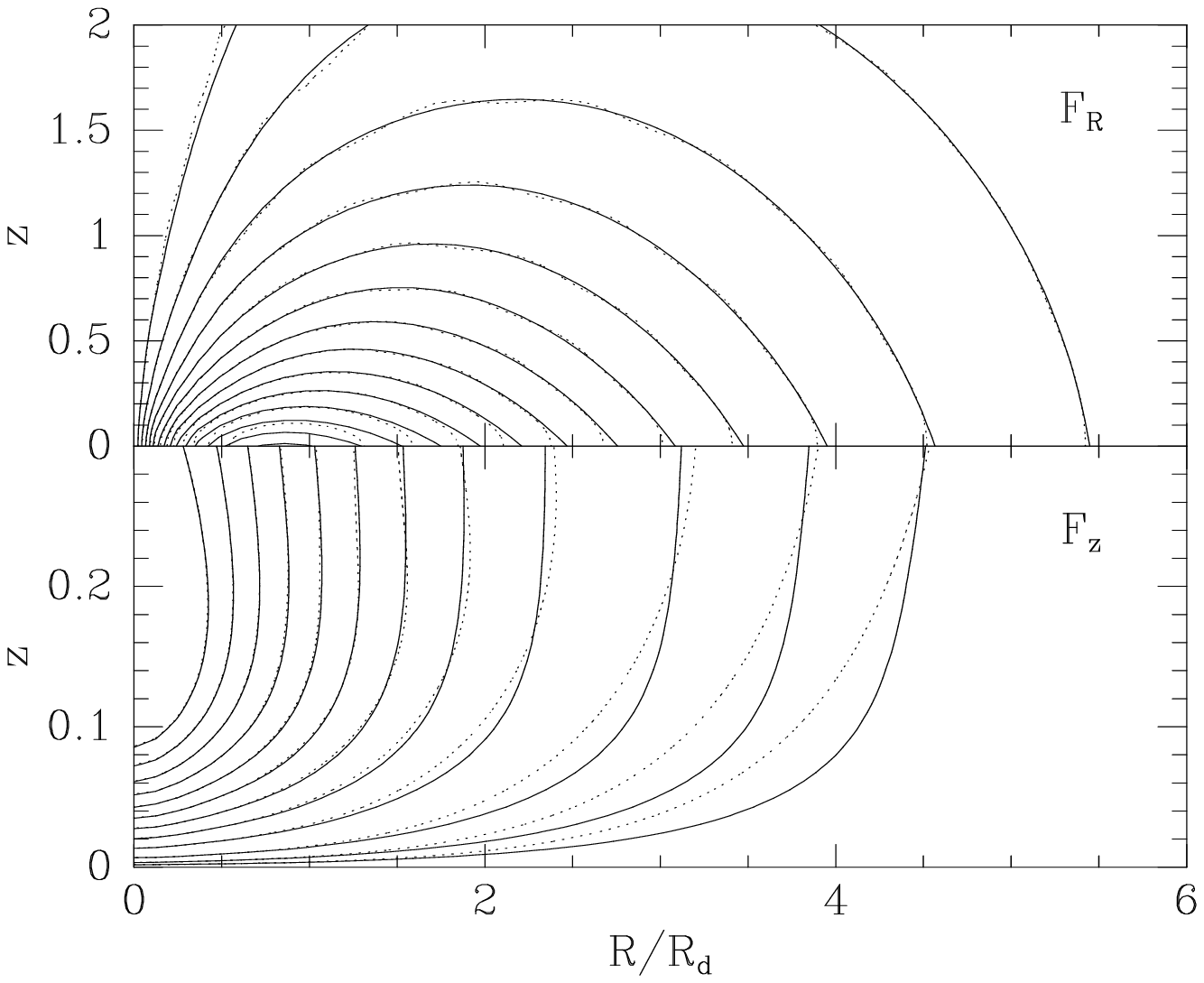,height=3.5in}}}
\caption{Representative contours of the resulting radial and vertical forces
for a $l=32$ pure multipole expansion ({\em dotted line})
and $l=8$ expansion of the residue plus $\Psi_{disk}$ ({\em solid line}).
Higher order expansions of the residue plus $\Psi_{disk}$
give the same result as the $l=8$ expansion.
The forces derived from the $l=32$ pure multipole expansion
still have significant errors at small $z$ and
intermediate $R$.}
\label{fig-frfz}
\end{figure}

Experience shows that, for a disk scale height of $0.15R_d$ (i.e.,
exponential scale height $R_d/13.3$ far from the disk plane) the
residual density can be well-fitted with a series truncated at $l=8$,
whereas even a series expansion to $l=32$ for the disk density alone
does not give satisfactory results (Figure \ref{fig-frfz}).

\section{A Sample Model}
\label{sect-nbody}
\noindent

The advantage of galaxy models built from distribution functions is
that the kinematics are fully specified. In this section we describe
how this kinematic information can be used to set up equilibrium
initial conditions for N-body simulations of disk-bulge-halo galaxies,
and provide some sample results.

\subsection{Setting up an N-body Realization}

We can generate an N-body realization of a galaxy by randomly sampling
from the \df's for each component.  The bulge and the halo are
straightforward to generate since the systems are nearly spherical and
the velocity ellipsoids are nearly isotropic.  The methods are
described in detail in Kuijken \& Dubinski (1994).  A particle's
position is first determined by sampling from the density
distribution.  With this position, one can find the local maximum of
the \df\ (at $(v_x,v_y,v_z)=(0,0,0)$) and then use the
acceptance-rejection technique to find a velocity.  This involves
selecting the three components of the velocity at random from a
velocity sphere with radius equal to the escape velocity.  A random
value, $f_{ran}$, of the \df\ is also chosen between 0 and the local
maximum.  If $f_{ran}$ is less than the value of the \df\ at the
chosen velocity then the velocity is accepted, otherwise it is
rejected and another attempt is made.

Sampling from the disk \df\ is slightly more troublesome, since the
disk is thin and the local velocity maximum of the disk \df\ must be
found at each point (it is not at ${\bf v}=0$ as for the bulge and
halo).  As before, we sample first from the density distribution to
find particle positions and then from the \df\ to find velocities.
Since the disks are generally warm, the maximum of the velocity
distribution occurs at a point where the azimuthal component of
velocity $v_\phi$ is less than the local circular velocity,
$v_{circ}$.  $v_R$ and $v_z$ are zero at the local maximum.  We find
the local velocity maximum for each particle position using standard
methods (Press et al. 1993) and then use the acceptance-rejection
technique as before to select a velocity.

\subsection{An N-body Realization}

\begin{figure}
{\centerline{\psfig{figure=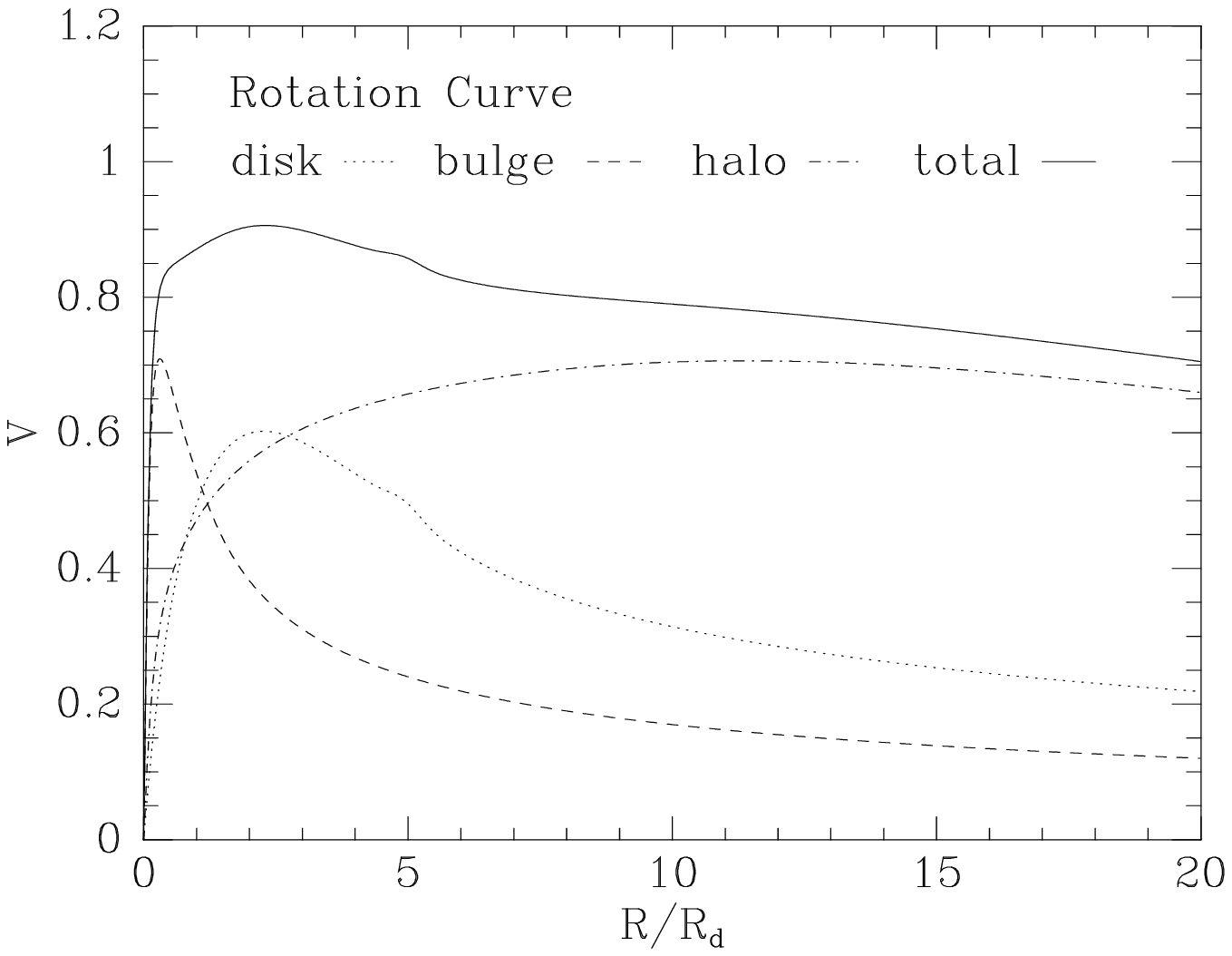,height=3.5in}}}
\caption{Rotation curve of a sample model.}
\label{fig-vr}
\end{figure}

We set up an N-body realization of a sample model to test the
validity of the equilibrium.  This galaxy model has
disk:bulge mass ratio of 4:1 and a halo:disk mass ratio of 1.5:1 within 5
disk scale lengths.  The total mass of the halo is about 10 times the disk
mass with the halo extending to 40 disk scale lengths (see Table 1 and 2
for the model parameters and resulting properties).
The rotation curve is fairly flat out to a radius of 10 disk scale lengths
(Figure \ref{fig-vr}).

\begin{table*}
\begin{minipage}{6.5in}
\caption{Galaxy Model Parameters. }
\begin{tabular}{rrrrrrrrrrrrrrrrr}
&~& \multicolumn{5}{c}{\sc Disk} &~&  \multicolumn{3}{c}{\sc Bulge} &~&
\multicolumn{5}{c}{\sc Halo} \\
Model &~&  $M_d$ &  $R_d$ & $R_t$ & $z_d$ & $\delta R$ &~&
$\Psi_{c}$ & $\sigma_b$ & $\rho_b$ &~&
$\Psi_o$ & $\sigma_0$ & $q$ &  $C$ & $R_a$ \\
&~& (1) & (2) & (3) & (4) & (5) &~&
(6) & (7) & (8) &~&
(9) & (10) & (11) & (12) & (13) \\ \hline
Sample &~& 1.00 & 1.0 & 5.0 & 0.15 & 0.3 &~&
-2.0 & 0.50 & 10.0 &~&
-4.0 & 1.00  & 0.9  & 0.1  & 0.5 \\
MW--A &~& 0.87 & 1.0 & 5.0 & 0.10 & 0.5 &~&
-2.3 & 0.71 & 14.5 &~&
-4.6 & 1.00  & 1.0  & 0.1  & 0.8 \\
MW--B &~& 0.87 & 1.0 & 5.0 & 0.10 & 0.5 &~&
-2.9 & 0.71 & 14.5 &~&
-5.2 &  0.96 &  1.0 &  0.1  & 0.8  \\
MW--C &~& 0.87 & 1.0 & 5.0 & 0.10 & 0.5 &~&
-3.7 & 0.71 & 14.5 &~&
-6.0 & 0.93 & 1.0 & 0.1 & 0.8 \\
MW--D &~& 0.87 & 1.0 & 5.0 & 0.10 & 0.5 &~&
-4.7 & 0.71 & 14.5 &~&
-7.0 & 0.92 & 1.0 & 0.1 & 0.8 \\
\hline
\end{tabular}
\medskip

(1) disk mass, (2) disk scale radius, (3) disk truncation radius,
(4) disk scale height, (5) disk truncation width, (6) bulge cut off potential,
(7) bulge velocity dispersion, (8) bulge central density ,
(9) halo central potential
(10) halo velocity dispersion,
(11) halo potential flattening, (12) halo concentration, $C=R_c^2/R_K^2$,
(13) characteristic halo radius.
\end{minipage}
\end {table*}

\begin{table*}
\begin{minipage}{6.5in}
\caption{Galaxy Model Properties.}
\begin{tabular}{rrrrrrrrrrrrrrrrr}
&~& \multicolumn{3}{c}{\sc Disk} &~&  \multicolumn{2}{c}{\sc Bulge} &~&
\multicolumn{2}{c}{\sc Halo} \\
Model &~&  $M$ & $\sigma_{r,0}$ & $R_e/R_d $ &~&
$M$ & $R_e/R_d$ &~&
$M$ & $R_e/R_d$ \\
&~& (1) & (2) & (3) &~&
(4) & (5) &~&
(6) & (7) \\ \hline
Sample &~& 0.94 & 0.50 & 5.6 &~&
0.29 & ~1.7 &~&
9.6 & 44.9 \\
MW-A &~& 0.82 & 0.47 & 6.0 &~&
0.42 & ~1.0 &~&
{}~5.2 & 21.8 \\
MW-B &~& 0.82 & 0.47 & 6.0 &~&
0.43 & ~1.0 &~&
{}~9.6 & 30.1 \\
MW-C &~& 0.82 & 0.47 & 6.0 &~&
0.43 & ~1.0 &~&
19.8 & 44.0 \\
MW-D &~& 0.82 & 0.47 & 6.0 &~&
0.43 & ~1.0 &~&
37.0 & 72.8 \\
\hline
\end{tabular}
\medskip

(1) disk mass,
(2) disk central radial velocity dispersion, $\sigma_r,0$,
(3) disk radial extent (radius where density drops to zero),
(4) bulge mass,
(5) bulge radial extent,
(6) halo mass,
(7) halo radial extent.
\end{minipage}
\end {table*}

\begin{figure}
{\centerline{\psfig{figure=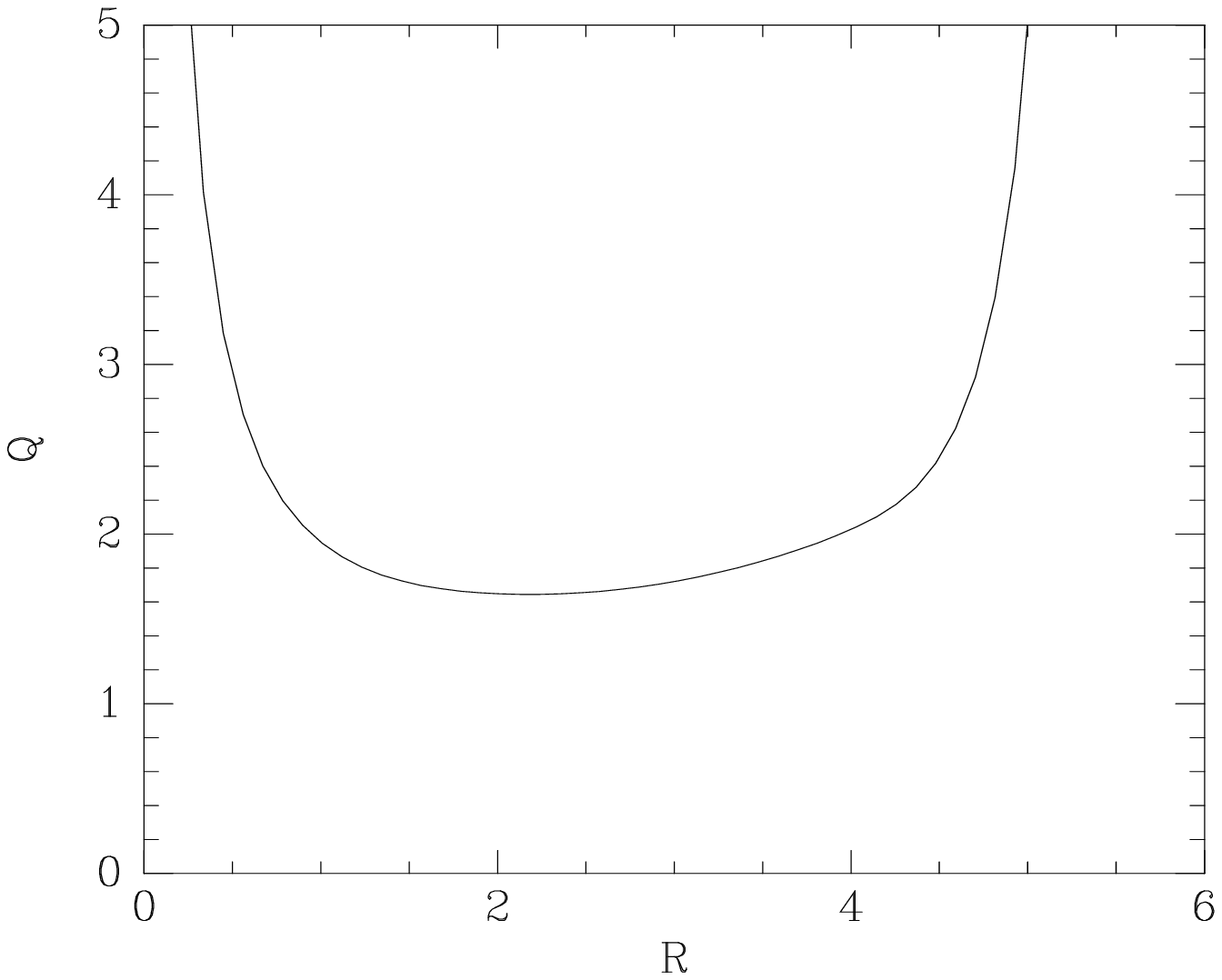,height=3.5in}}}
\caption{Toomre's stability parameter
$Q=\sigma_r \kappa/3.36 G\Sigma$ versus the radius}
\label{fig-toomre}
\end{figure}

The disk is warm with a Toomre $Q=1.7$ at the disk half mass radius.
The value of Q is fairly constant throughout the disk, though
rising both in the
center within 1 scale length and near the edge where the
surface density tapers off to zero (Figure \ref{fig-toomre}).

We performed various simulations to assess the validity of our approximate
disk \df\ and the stability of the overall system.
We used a tree code for all of the simulations (Barnes \& Hut 1986;
Hernquist 1987; Dubinski 1988).
The critical opening angle was set to $\theta = 0.9$ with forces between
cells and particles calculated to quadrupole order.  The particle softening
radius was set to 0.025 disk scale lengths (or 0.17 disk scale heights).
The orbital time of the model at the disk half mass radius is 13 units.
We set the leapfrog timestep to 0.1 units and ran the models to $t=96$ or
approximately 7.5 orbital times.
We set up 4 simulations:
\begin{enumerate}
\item  a disk of 40000 test particles orbiting in the
derived system potential;
\item  a gravitating disk of 40000 particles
with a static bulge and halo potential;
\item  a disk, bulge, and halo with
40000, 10000, and 50000 particles respectively; and
\item  a disk, bulge and
halo with 80000, 20000, and 200000 particles.
\end{enumerate}
For all of the models, we followed the time evolution of the disk
surface density profile, $\Sigma(R)$, the disk velocity dispersion
profile averaged in rings, and the disk scale height versus radius.
For the bulge and halo, we calculated the spherically averaged density
profiles.

\begin{figure*}
{\centerline{\psfig{figure=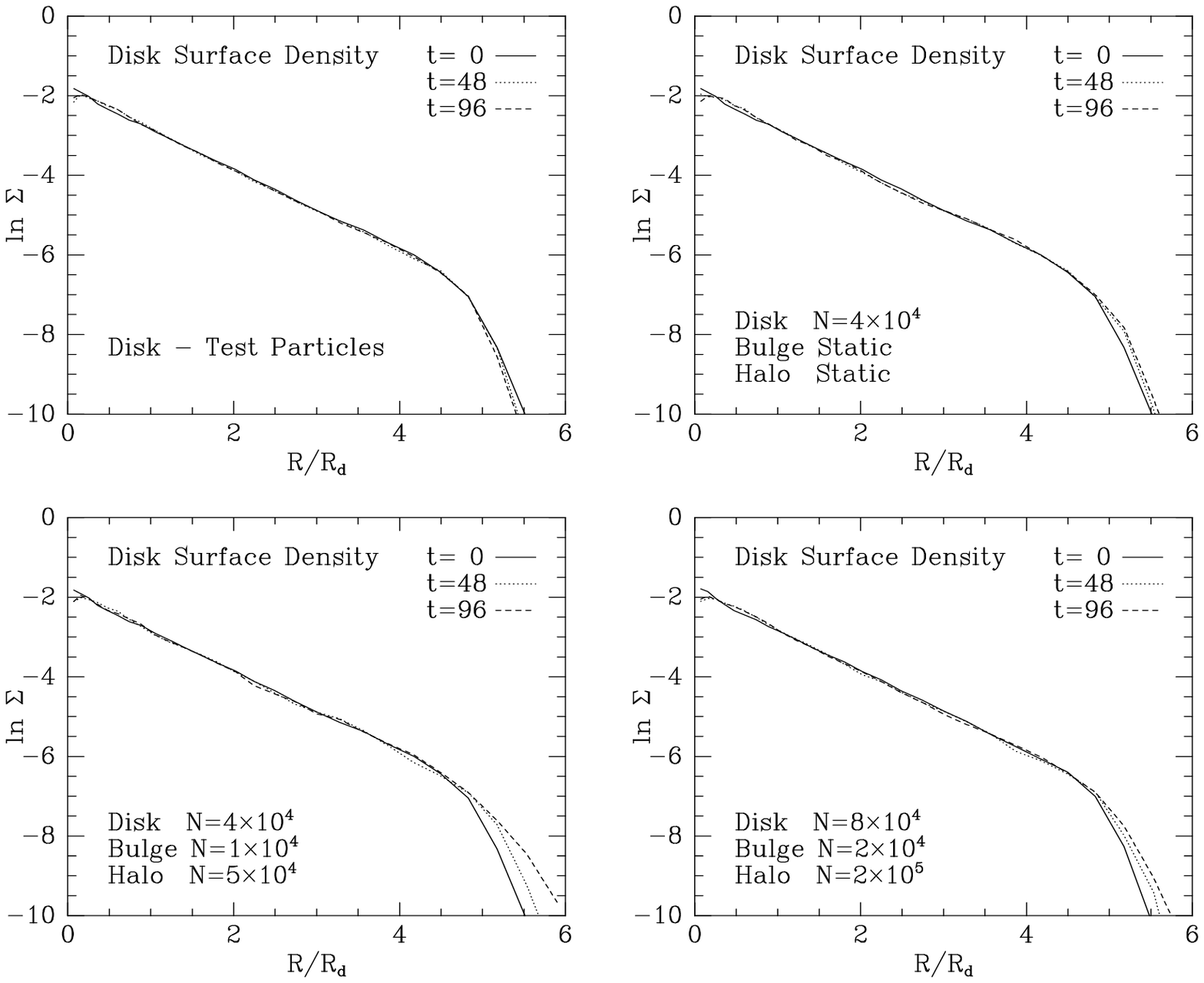}}}
\caption{The disk surface density profile as a function of time for the 4
simulations as labelled.}
\label{fig-dden}
\end{figure*}

\begin{figure*}
{\centerline{\psfig{figure=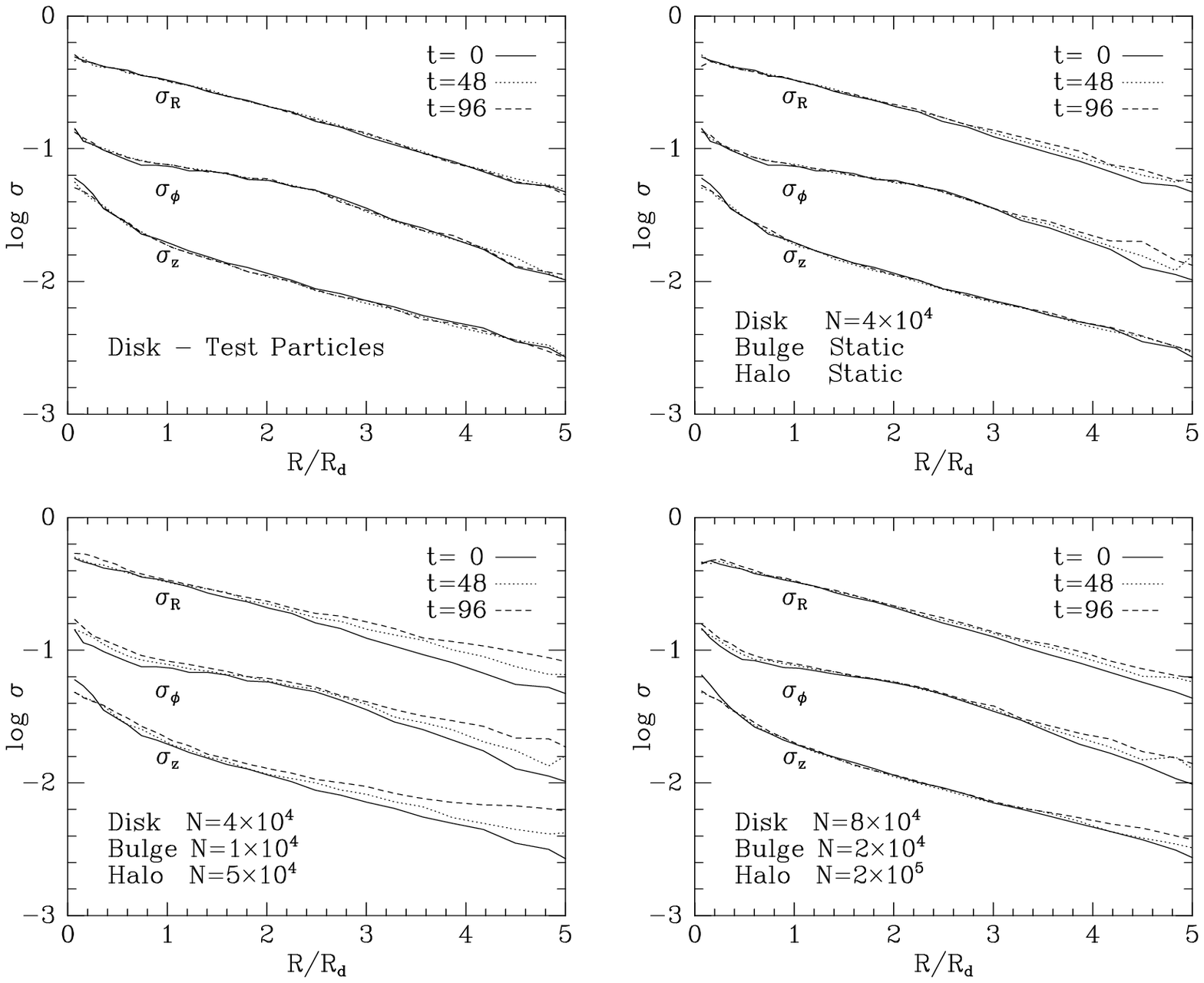}}}
\caption{The velocity dispersion profiles as a function of time for the 4
simulations as labelled.  To avoid overlap, $\sigma_\phi$ and $\sigma_z$ are
offset by 0.5 and 1.0 dex}
\label{fig-vsig}
\end{figure*}

The first simulation provides a simple check of our \df, since the
test particles should maintain a constant density and velocity
dispersion profile in the total potential of the model.  This test is
most critical for the disk since we have approximated a third integral
with the vertical energy, $E_z$ and it is not guaranteed that this
quantity is conserved sufficiently well to maintain a decent equilibrium.
For our sample model, we find that the approximation works very well.
There is essentially no change in the density profile (Figure
\ref{fig-dden}) and velocity dispersion profiles (Figure \ref{fig-vsig}).

\begin{figure*}
{\centerline{\psfig{figure=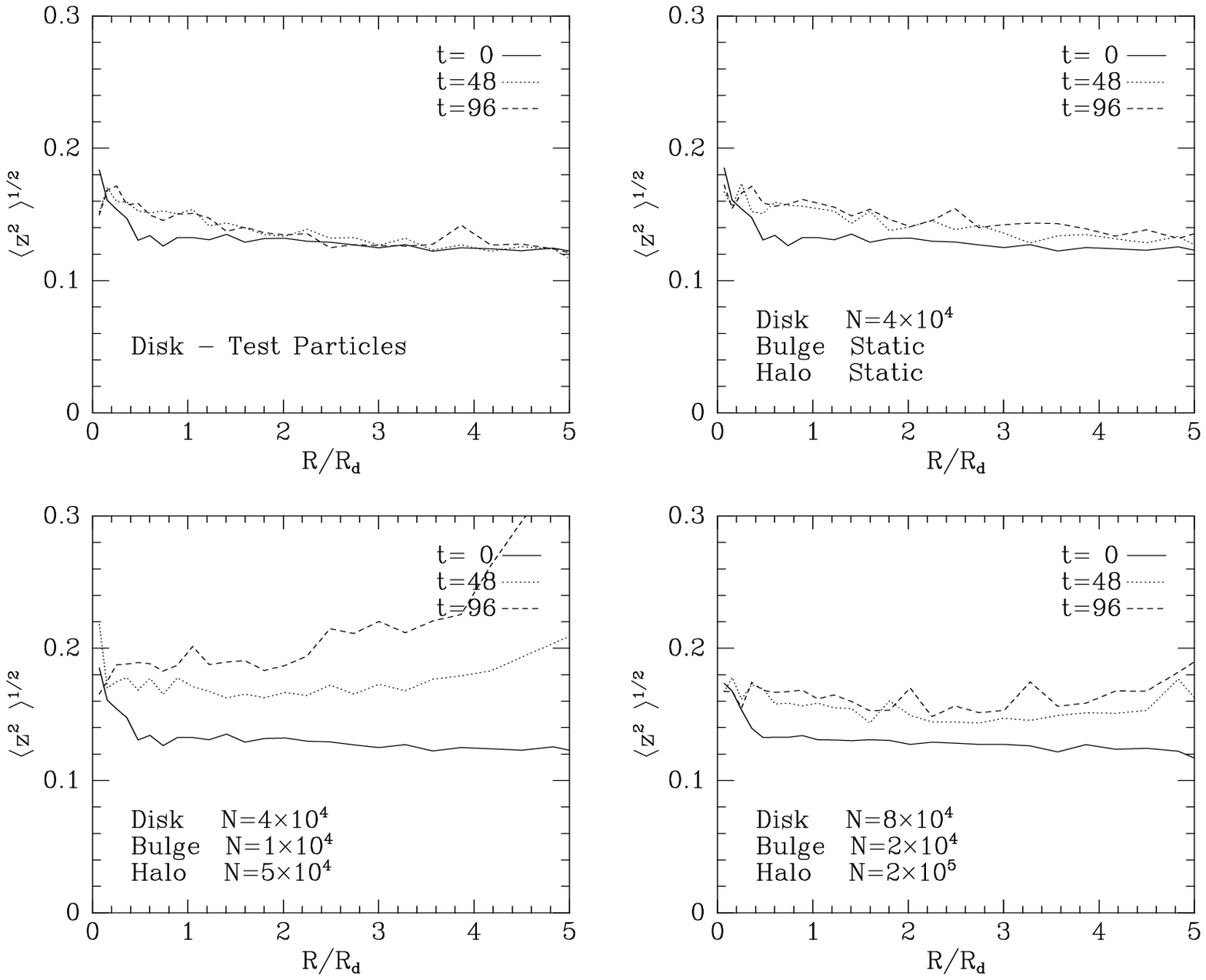}}}
\caption{The disk scale height $\langle z^2 \rangle^{1/2}$ as a function of
time for the 4 simulations as labelled.}
\label{fig-zh}
\end{figure*}

In the second simulation, we made the disk `live' while maintaining a
static bulge and halo potential.  We thereby avoid external
perturbations from a noisy bulge and halo potential (which we will see
below can significantly heat the disk).  The surface density profile
is nearly constant, once again deviating only slightly from the
initial conditions.  The disk is initially cool enough to suffer from
spiral arm instabilities and this mainfests itself as a gradual rise in
$\sigma_R$ and $\sigma_\phi$ throughout the disk
(Figure \ref{fig-vsig}) (e.g. Sellwood \& Carlberg 1984, Toomre \&
Kalnajs 1991).  This self-heating is unavoidable in cool disks, even
when set up from a \df\ which is formally in equilibrium.
The vertical velocity dispersion, $\sigma_z$, remains fairly constant
except at the edge of the disk where it is poorly sampled by particles
and discreteness noise is large, and near the center where the disk is
hot and the assumed constancy of $E_z$ along orbits built in to the \df\
breaks down.  Nevertheless, $\sigma_z$ does not show
any significant initial transients over most of the disk suggesting
that the disk \df\ represents a starting point very close to
equilibrium.  We do not expect spiral arm formation to heat the disk
vertically, since the modes grow in the disk plane and couple weakly
to vertical oscillations (e.g., Jenkins \& Binney 1990).
We also plotted the
r.m.s.\ height of disk particles averaged in rings (Figure \ref{fig-zh}).
There is a slight increase with time in the average scale height which may
arise from either the particle softening, which would
weaken the local disk self-gravity and allow it to puff up,
or from gradual heating by the disk's own discreteness.

The final two simulations, in which the halo and bulge are also live,
show the strong effect of a coarse-grained halo on the disk evolution.
A particle halo adds a significant source of disk heating (Hernquist
1993) when the number of halo particles is too small.  The halo is
effectively composed of massive black holes that bombard the disk as
in the model of Lacey \& Ostriker (1985).  The continuous bombardment
of halo particles heats the disk at a rate $\sigma^2 = (\sigma_0^2 + D
t)$, where $D$ is proportional to the black hole mass and local halo
density.  The disk velocity ellipsoid therefore changes and the
vertical scale height increases.  This effect is strongest at the disk
center where the halo density is the greatest, and is weakened as
the number of halo particles is increased, in agreement with the
Lacey-Ostriker results.  The graininess of the halo can also excite
bending instabilities especially in the outer disk where the disk
density is low.  The disks in simulation 3 do indeed exhibit bending
instabilities that resemble a flag waving in the breeze.  This
behaviour was less apparent in the larger simulation 4.  Again as in
simulation 2 gravitational softening can also cause the disk to puff
up.

Figure
\ref{fig-vsig} shows the velocity dispersion profiles for the
simulations with particle bulges and halos.  We can see that there is
significant heating for the smaller N model over and above the heating
of the disk in the static halo potential.  The disk heating for the
large N halo is significantly smaller with the $\sigma_z$ increasing
only at disk radii with $R > 3 R_d$.  The disk scale height also does
not grow as much for the larger N model in accord with expected
heating rates (Figure \ref{fig-zh}).
The models are apparently converging to a smooth equilibrium
limit as we increase $N$.

\begin{figure}
{\centerline{\psfig{figure=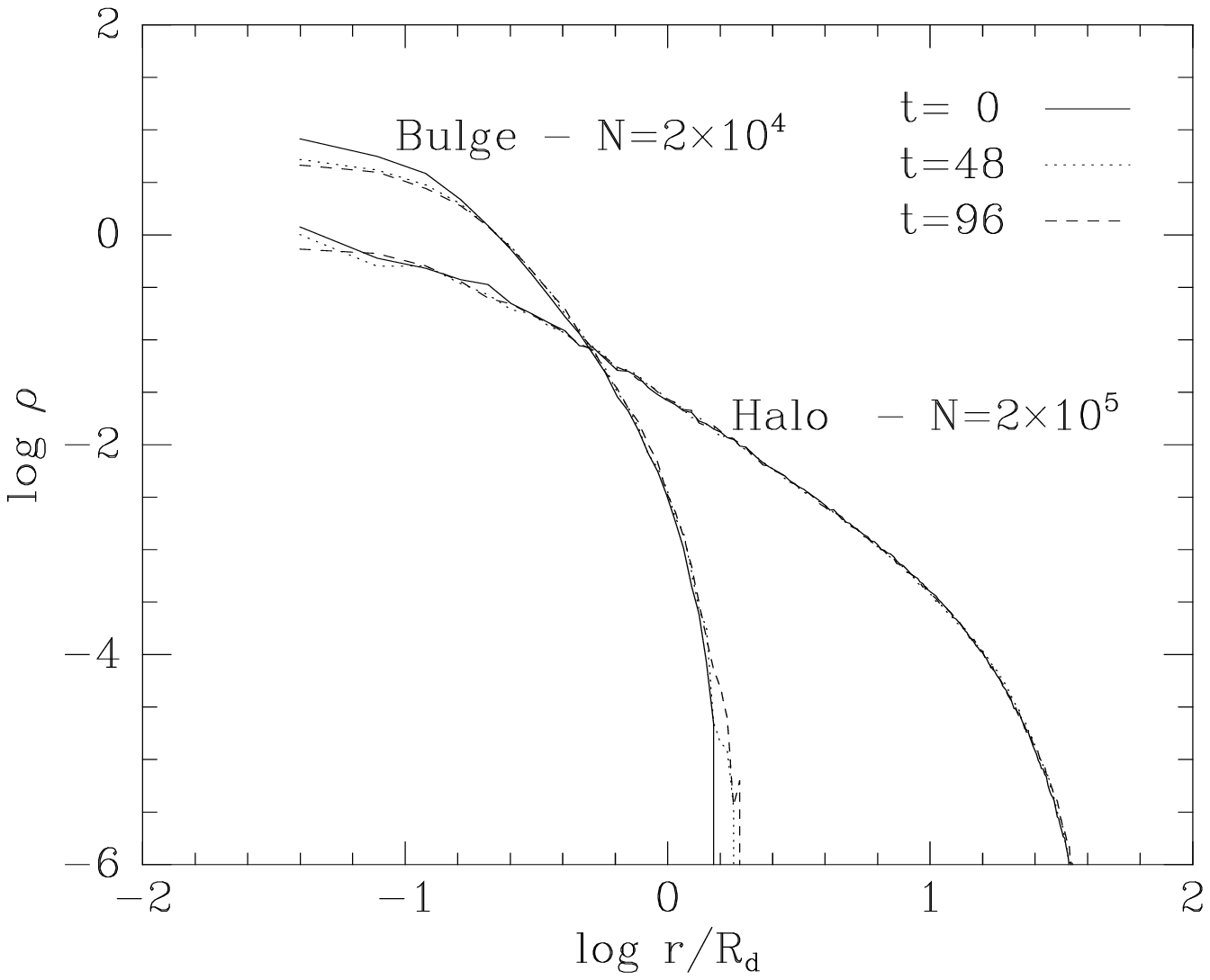,height=3.5in}}}
\caption{The bulge and halo density profiles for largest simulation as a
function of time.  The bulge relaxes to a slightly lower density for
$r<0.1$ probably as a result of the gravitational softening,
but overall the density profiles remain constant.}
\label{fig-bhden}
\end{figure}

Unlike the disk, the bulge and halo maintain their integrity
during the simulations, as expected since they are derived
from \df's with exact integrals of motion in the given potential.
Figure \ref{fig-bhden} shows the constancy of the
density profiles of the models at 3 times.
The slight drop in the density in the bulge within $r < 0.1$
can be attributed to the particle softening radius $\epsilon = 0.025$.
The reduced gravity from the softening allows the bulge to expand a little
at the center and therefore drop in density.
Despite this minor defect, the bulge and halo stay in equilibrium.

At late times (5-8 orbital times), a modest bar begins to develop
within 1 scale radius in the completely live models.  The bar forms
sooner in the smaller $N$ models.  The seeds of these bar
instabilities probably originate in the discreteness noise of the halo
also observed by Hernquist
(private communication).  The onset of bar formation can
only be delayed by increasing $N$.

In summary, a disk \df\ with $E_z$ as an approximate third
integral results in a disk-bulge-halo galaxy model very close to formal
equilibrium.  We can therefore follow subtle changes in the disk structure
and kinematics resulting from a variety of instabilities.

In the next section, we find a set of model parameters which produce a mass
model closely resembling the Galaxy.

\section{Models of the Milky Way}
\label{sect-mw}

\begin{figure}
{\centerline{\psfig{figure=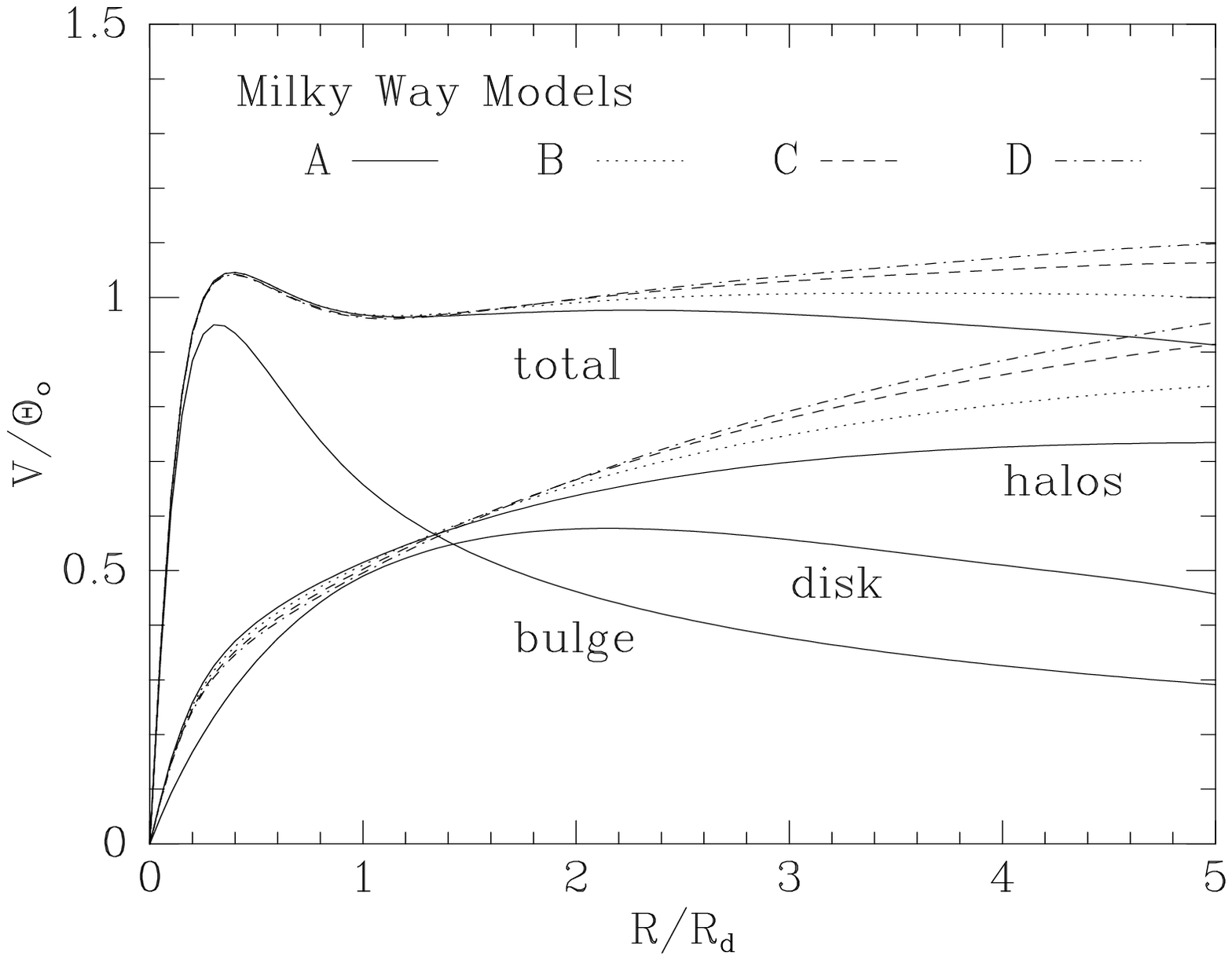,height=3.5in}}}
\caption{The rotation curves for the Milky Way models, MW-A, B, C, and D,
showing contributions from the disk, bulge and halo in the inner regions
within $R<5R_d$.
}
\label{fig-vr-mw-inner}
\end{figure}

\begin{figure}
{\centerline{\psfig{figure=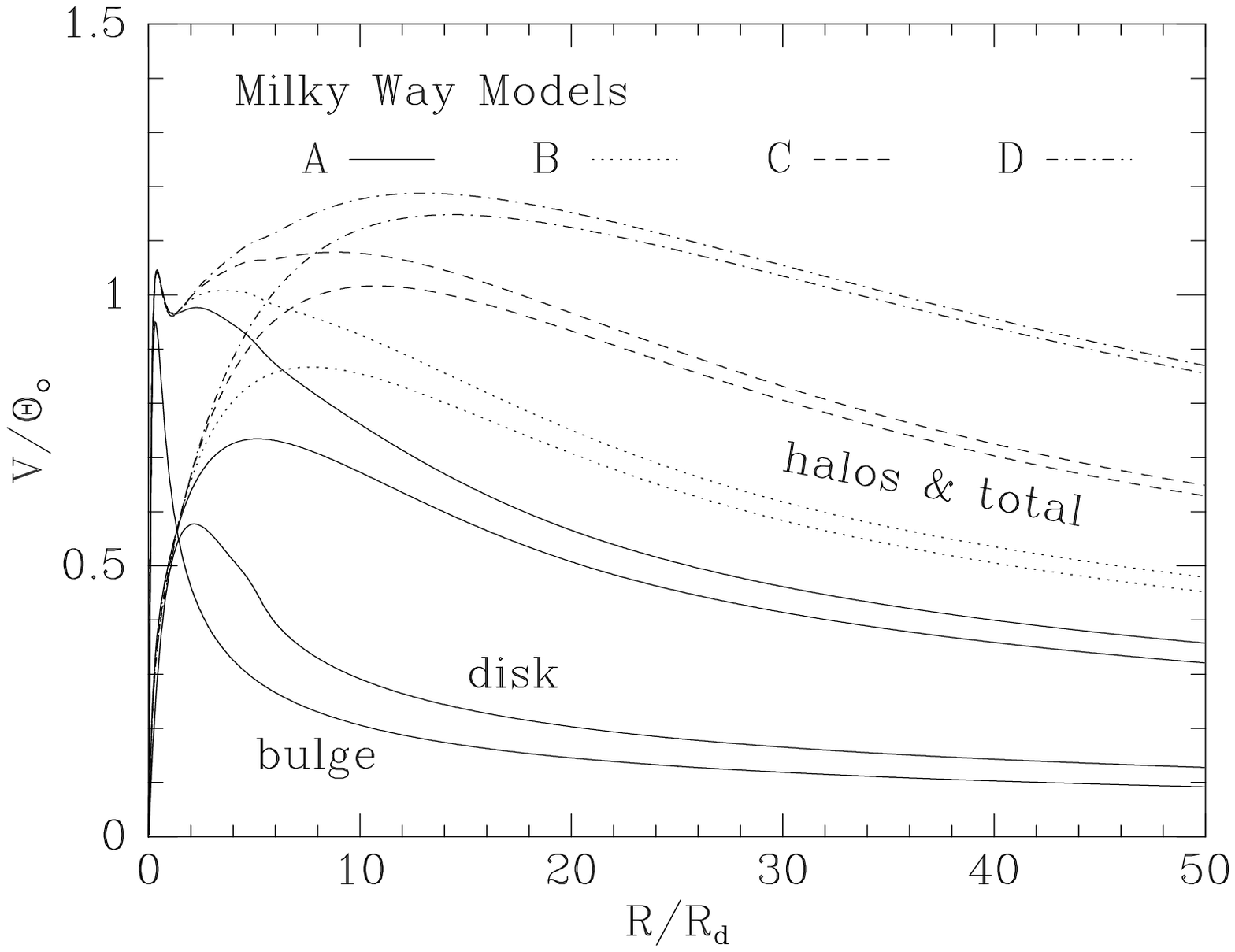,height=3.5in}}}
\caption{The rotation curves for the Milky Way models, MW-A, B, C, and D,
showing contributions from the disk, bulge and halo in the outer regions
out to $R=50R_d$.}
\label{fig-vr-mw-outer}
\end{figure}

In Table 1, we present the parameters for generating a sequence of 4
models, MW-A,B,C, and D, which have mass distributions and rotation
curves closely resembling those of the Milky Way within 5 scale radii.
The disk and bulge mass distributions are the same for each model with
mass and extent of the halo increasing through the sequence (Table 2).
The halos are all chosen with $q=1.0$, though they are slightly
squashed in the self-consistent galaxy models.  Model MW-D has the
halo with the largest mass and has the most realistic representation
of the outer galaxy.  These models were found by trial and error and
renormalized so that the flat portion of the rotation curve had $V_c
\approx 1.0$. The contributions to the radial acceleration in the
solar neighbourhood ($R=1.8R_d$) from disk, bulge and halo are
comparable in these models, as found by Kuijken \& Gilmore (1989) in
their study of the local disk surface density.  The natural units for
length, velocity, and mass for these dimensionless models are $R_d =
4.5$ kpc, $V=220$ km s$^{-1}$, and $M=5.1 \times 10^{10}$ M$_\odot$.
The central velocity dispersion was chosen so that the observed radial
velocity dispersion of 42 km s$^{-1}$ at the solar radius ($R=1.8
R_d$) would be reproduced in the model.  Figures \ref{fig-vr-mw-inner}
and \ref{fig-vr-mw-outer} show the rotation curves out to 5 scale
radii and 50 scale radii respectively.

We tested the stability of the models against bar formation and found that
they were not immediately unstable.
A small bar within 1 scale radius did form after four orbital times, though
increasing the number of particles delayed the formation to later times.
It seems that the observed instability was due to the discreteness
of the simulation rather than the intrinsic mass distribution.
In the limit of large N, these particular models are probably stable
against bar formation.
The fact that we observe a bar in the galactic center
(e.g. Weinberg 1992) might imply a larger disk:halo mass radius within
the solar circle than these models assume.

\section{Summary}
\label{sect-discuss}

We have described a method for setting up self-consistent galaxy
models with a disk, bulge and halo where each component is described
by a separate phase-space distribution function (\df).  The halo \df\
depends on particle energy $E$ and angular momentum $L_z$, and has the
form of a lowered Evans model (Kuijken \& Dubinski 1994), resulting in
a flattened, finite radius model. The bulge \df\ is a King model (King
1966) which is a function only of $E$. We introduce a new \df\ for the
disk, a version of the planar \df\ of Shu~(1969) modified to include
vertical structure. It is a function of $E$, $L_z$ and $E_z$, the
vertical energy.  The disk \df\ is approximate in that the third
``integral'', $E_z$, introduced to describe the vertical motion, is
not conserved exactly.  Nevertheless, in practice the disk \df\ yields
a model very close to equilibrium. The three components respond to the
total gravitational potential of the model. The main advantages of
these models over previous ones are:
\begin{enumerate}
\item The flattening and rotation of the halo can be specified in the
Lowered Evans model \df, which allows examination of new problems
where these parameters are important such as the dynamics of galactic
warps and satellite orbital evolution.
\item The disk does not suffer from additional transient adjustments
at start up beyond the expected (and unavoidable) spiral and bar
instabilities.  The potential of the system and velocity ellipsoids
(particularly those of the disk)
will therefore not change
significantly from the initial state.  These models are therefore
useful for studying problems involving the effects of small
perturbations on the disk such as disk heating by sinking satellites
or external tidal fields.
\item Since the \df\ is known explicitly, many quantities (line
profiles, for example) can be calculated directly without the need for
simulation.
\end{enumerate}

These same techniques can be generalized to models with different
bulge and halo \df's and disks with different density distributions.
Of course, equilibrium does not imply stability, and it is by no means
guaranteed that these models are free from spiral and bar
instabilities.  In practice, though, models can be found which are
apparently stable stable over many rotation periods.  Disk heating by
the halo is troublesome, and ultimately can only be avoided by going
to very large $N$ ($\sim 10^6$) if we wish to evolve a system for a
Hubble time.

\end{document}